\documentclass[aps,prl,twocolumn,superscriptaddress]{revtex4}
\usepackage{amsmath}
\usepackage{graphicx}
\usepackage{dcolumn}
\usepackage{bm}
\usepackage{amsfonts}
\usepackage[dvips]{color}

\makeatletter
\def\captionof#1#2{{\def\@captype{#1}#2}}
\makeatother

\newcommand{\rme}{{\rm e}}
\newcommand{\rmi}{{\rm i}}
\newcommand{\ud}[1]{\hspace{-0em}\mathrm{d}{#1}\;}
\newcommand{\Ud}[1]{\hspace{-0.5ex}\mathrm{d}{#1}\;}
\newcommand{\udd}[2]{\hspace{-0em}\mathrm{d}{#1}\,\mathrm{d}{#2}\;}

\begin{document}
\title{Impurity Model for Non-Equilibrium Steady States}

\author{Camille Aron}
\affiliation{Department of Physics and Astronomy, Rutgers University, 136 Frelinghuysen Rd., Piscataway, NJ 08854, USA}
\author{Cedric Weber}
\affiliation{King's College London, Theory and Simulation of Condensed Matter (TSCM), The Strand, London WC2R 2LS}
\author{Gabriel Kotliar}
\affiliation{Department of Physics and Astronomy, Rutgers University, 136 Frelinghuysen Rd., Piscataway, NJ 08854, USA}

\begin{abstract}
 We propose an out-of-equilibrium impurity model for the dynamical mean-field description of the Hubbard model driven by a finite electric field. The out-of-equilibrium impurity environment is represented by a collection of \emph{equilibrium} reservoirs at different chemical potentials. We discuss the validity of the impurity model and propose a non-perturbative method, based on a quantum Monte Carlo solver, which provides the steady-state solutions of the impurity and original lattice problems. We discuss the relevance of this approach to other non-equilibrium steady-state contexts.

\end{abstract}

\maketitle

The understanding and techniques of condensed matter theory are such that it is possible nowadays to make reliable quantitative predictions on equilibrium properties of many materials, including strongly-correlated many-body systems. Despite significant progress, in comparison very little is know about correlated systems far from equilibrium. Even for simple measurements such as current-voltage characteristics, one can only hope for some qualitative theoretical understanding. In this perspective, the study of non-equilibrium steady states (NESS) is paramount for understanding which of the equilibrium concepts, techniques and results can be adapted to non-equilibrium situations.

The adaptation of equilibrium numerical methods to time-dependent situations (\textit{e.g.} exact diagonalization, density matrix renormalization group, diagrammatic Monte Carlo, dynamical mean-field theory, variational methods) enable the study of complex transients such as the dynamics after a quench in temperature or interaction.
However, the limitations in time of such time-dependent methods (\textit{e.g.} size or Poincare recurrence effects for closed systems, growth of entanglement entropy, sign problem) usually make them unsuitable for the study of steady-state physics.

Recent theoretical efforts have been done to address the non-equilibrium steady-state physics of quantum dots driven by a constant voltage by developing steady-state formalisms that bypass the transient dynamics~\cite{Natan, Karyn}. Among several promising steady-state techniques~\cite{NESS-Algo-dot}, a remarkable leap forward has been achieved recently by Han and Heary who developed a non-equilibrium steady-state impurity solver based on a Matsubara-like formalism and a Hirsch-Fye quantum Monte Carlo algorithm (NESS-QMC)~\cite{Han-Heary, Han}.
Progress has also been made in describing the non-equilibrium steady-state dynamics of correlated electrons on finite dimensional lattices~\cite{Onoda, Knap}. In the context of correlated systems driven by electric fields, non-perturbative steady-state solutions have been computed by adapting the dynamical mean-field theory to these non-equilibrium steady states (NESS-DMFT)~\cite{NESS-DMFT, AronKotliarWeber2012}.

In this work, we give substance to the NESS physics of such lattice problems by focusing on their local mean-field description (the so called impurity problem), establishing the connection with the out-of-equilibrium physics of quantum dots. In the framework of the electric-field-driven Hubbard model, we model the environment of the related impurity problem by a collection of \emph{equilibrium} leads at different chemical potentials. We show how non-equilibrium properties on the lattice side, such as the energy distribution function or the dissipation, translate on the impurity side. Simplifying the impurity model by truncating it to the first relevant leads, we solve it numerically \textit{via} a generalized NESS-QMC solver \textit{\`a la} Han-Heary, and we obtain the corresponding NESS-DMFT solution of the electric-field-driven Hubbard model. We compare the results with the ones obtained by solving the impurity to the second order in the interaction with the so-called iterated perturbation theory (IPT).

\medskip

\paragraph{\textbf{Lattice model}.} 
We consider the Hubbard model on a $d=2$ square lattice. It is driven out of equilibrium by a static and uniform electric field set along the $x$-axis of the lattice: $\mathbf{E} = E \mathbf{u}_x$ with $E>0$. The corresponding Lagrangian reads (we set $\hbar=1$)
\vspace{-0.5em}
\begin{equation}
\begin{array}{rl}\label{eq:Lagrangian}
\mathcal{L}_{s} =& \displaystyle \sum_{i \sigma} \bar c_{i \sigma} \left[ \rmi\partial_t -  \phi_{i}(t) \right] c_{i \sigma} 
 - U \sum_{i}  \bar c_{i \uparrow}  c_{i  \uparrow}  \bar c_{i \downarrow}  c_{i \downarrow} \\ 
  & +  \displaystyle \sum_{\langle i j \rangle \sigma}   \bar c_{i \sigma} t_{ij} \rme^{\rmi\alpha_{ij}(t)} c_{j \sigma} 
+ \mathrm{conj.} 
\end{array}
\end{equation}
where $c_{i\sigma}$ and $\bar c_{i\sigma}$ are the Grassmann fields representing an electron at site $i$ with spin $\sigma\in\{\uparrow,\downarrow\}$. The Hubbard $U$ term accounts for the on-site Coulombic interaction and $t_{ij} \equiv (a/2\pi)^{2} \int\Ud{\mathbf{k}} \rme^{\rmi\mathbf{k}\cdot\mathbf{x}_{ij}} \epsilon(\mathbf{k}) $ is the hopping amplitude between two nearest neighbors distant of $a$: $\epsilon(\mathbf{k}) = \epsilon_0 \left[\cos(k_x a)+\cos(k_y a) \right]$.
$\alpha_{ij}(t) \equiv q \int_{\mathbf{x}_j}^{\mathbf{x}_i} \Ud{\mathbf{x}}\!\cdot\!\mathbf{A}(t,\mathbf{x})$ are the Peierls phase factors,  $q$ is the charge of the electrons and $\mathbf{A}$ ($\phi$) is the vector (scalar) potential: $\mathbf{E} = -\boldsymbol{\nabla} \phi - \partial_t \mathbf{A}$.

To allow for non-trivial steady states, we couple the system to a thermostat composed of independent reservoirs in equilibrium at temperature $T$: ${\cal L}_{sb} = \gamma \sum_{i \sigma l} \rme^{\rmi\theta_i(t)} \bar b_{i\sigma l} c_{i\sigma} + \mbox{conj.}\,$~\cite{Adriano}. The $b_{i\sigma l}$'s are non-interacting electrons, and $l$ labels their energy level. The phases $\theta_i(t) \equiv \int^t \Ud{t'} \phi_i(t')$ ensure the $\mathrm{U}(1)$ gauge invariance and we work with a particle-hole symmetric system by simply considering half-filled reservoirs with a flat density of states of bandwidth $W$. The thermostat introduces a local (gauge-invariant) retarded hybridization $\mbox{Im} \Sigma_{\rm th}^R(\omega)  = -\Gamma$ where $\Gamma \equiv \gamma^2/W$ sets the dissipation rate. The corresponding Keldysh component of the hybridization is fixed by the fluctuation-dissipation theorem: $\Sigma_{\rm th}^K= (2 f_T -1) \mbox{Im} \Sigma_{\rm th}^R$ where $f_{T}(\epsilon) \equiv [1+\exp(\epsilon/T)]^{-1} $ is 
the equilibrium Fermi-Dirac distribution at temperature $T$. We take this temperature to be the lowest energy scale ($k_B = 1$). We restrict ourselves to the paramagnetic solution and drop the spin indices. In the numerics, we use $q=a=1$ and measure energies in units of $\epsilon_0$.

For uniform non-equilibrium steady states, the Schwinger-Dyson equations of motion read~\cite{AronKotliarWeber2012}
\begin{align} \label{eq:DysonLattice}
\left\{
\begin{array}{l}
 \left[ \omega + \epsilon(\mathbf{k}) - \Sigma^R(\omega,\mathbf{k}) \right] \ast  G^R(\omega,\mathbf{k}) = 1\,, \\
  G^K(\omega,\mathbf{k}) = G^R(\omega,\mathbf{k}) \ast \Sigma^K(\omega,\mathbf{k}) \ast G^R(\omega,\mathbf{k})^*\,, \\
\end{array}
\right.
\end{align}
where  $G^{R/K}(\omega,\mathbf{k})$ and $\Sigma^{R/K}(\omega,\mathbf{k})$ are respectively the gauge-invariant 
retarded/Keldysh Green's functions and self-energies that correspond to $G^{R/K}({\varpi,\boldsymbol{\kappa}})$ and $\Sigma^{R/K}(\varpi,\boldsymbol{\kappa})$ in~\cite{AronKotliarWeber2012}.
Both the thermostat and the Coulombic interaction contribute to the self-energy kernels: $\Sigma^{R/K} = \Sigma_{\rm th}^{R/K}  + \Sigma_{U}^{R/K}$. The star product $\ast\equiv \exp{(\frac{\rmi}{2}q
[  \overleftarrow{\partial_\omega} \, \overrightarrow{\mathbf{E} \cdot\! \boldsymbol{\nabla}_{\!\mathbf{k}}}
\!- \overleftarrow{\mathbf{E} \cdot\!\boldsymbol{\nabla}_{\!\mathbf{k}}}\, \overrightarrow{\partial_\omega}  
])
}
$, in which left (right) arrows indicate derivative operators acting on the left (right),
is a consequence of working with gauge invariance quantities.

\medskip

\paragraph{\textbf{Dynamical mean-field approach}.} 
In equilibrium, the self-energy of a $d$-dimensional Hubbard model reduces to the one of a dimensionless Anderson Impurity Model in the limit of infinite connectivity. Dynamical mean-field theory (DMFT) uses this result to compute non-perturbative mean-field solutions of finite dimensional lattice problems by solving auxiliary impurity problems defined self-consistently~\cite{DMFT}.
Here, we propose to generalize this mapping for non-equilibrium steady states.

The Keldysh action of a single-band Hubbard-$U$ impurity in a generic steady state reads 
\begin{equation}\label{eq:Simp}
 \begin{array}{lcl}
S&=&\displaystyle\sum_{ab} \sum_{\sigma} \iint\udd{t}{t'} \bar c_\sigma^a(t) {\mathcal{G}_0^{-1}}^{ab}(t-t') c_{\sigma}^b(t') \\ & &\displaystyle - \sum_{a} a \int \ud{t} U    \bar c^a_{\uparrow}(t) c^a_{\uparrow}(t)  \bar c^a_{\downarrow}(t)  c^a_{\downarrow}(t)\,,
\end{array}
\vspace{-0.5em}
\end{equation}
where the indices $a, b = \pm$ refer to the forward and backward branch of the Keldysh contour.
The $\mathcal{G}_0^{ab}$ are the impurity non-interacting  Green's functions (often referred as the Weiss effective fields) that include the local hybridization with the thermostat. The two quantities that fully describe the non-equilibrium steady-state physics of the impurity are its density of states $\rho(\epsilon)$ and its energy distribution $f(\epsilon)$ (given by the Fermi-Dirac distribution in equilibrium). Both are encoded in the interacting Green's functions  through the relations
\begin{align}\left\{
\begin{array}{rl}
 {\mbox{Im}  {\cal G}^R(\omega)}  &= - \pi \, \rho(\omega) \,,\\
 {\cal G}^K(\omega) &= [2 f(\omega) -1] \, \mbox{Im}  {\cal G}^R(\omega)\,,
\end{array}
\right.
\end{align}
where  ${\cal G}^R = {\cal G}^{++} - {\cal G}^{+-} $  and $ {\cal G}^K = \rmi[{\cal G}^{++} + {\cal G}^{--}]/2$ are respectively the retarded and the Keldysh Green's functions of the impurity. They obey the following Schwinger-Dyson equations
\begin{align} \label{eq:SchwingerDyson}
\left\{
\begin{array}{rl}
 {\cal G}^R(\omega) &= \left[ {\cal G}_0^R(\omega)^{-1}-\Sigma_U^R(\omega) \right]^{-1}\,, \\
 {\cal G}^K(\omega) &= \left| {\cal G}^R(\omega) \right|^2 \left[  \frac{{\cal G}_0^K(\omega)}{|{\cal G}_0^R(\omega)|^2} + \Sigma_U^K(\omega) \right] \,,
\end{array}
\right.
\end{align}
where $\Sigma_U^{R/K}$ are the impurity self-energy kernels stemming from the local Coulombic interaction in Eq.~(\ref{eq:Simp}).
DMFT consists of approximating the lattice self-energy by the one of the auxiliary impurity problem: $\Sigma^{R/K}(\omega,\mathbf{k}) \simeq \Sigma^{R/K}(\omega)$.
In order for the impurity problem to locally describe the lattice problem, one imposes that it has the same (local) density of states and distribution function.
This corresponds to the following self-consistent equations for the interacting Green's functions
\begin{align} \label{eq:SelfConsistent}
\left\{
\begin{array}{rl}
 {\cal G}^R(\omega) &= G^R(\omega) \,,  \\
 {\cal G}^K(\omega) &=G^K(\omega)\,,
\end{array}
\right.
\end{align}
where $G^{R/K}(\omega) = (a/2\pi)^2 \int\ud{\mathbf{k}} G^{R/K}(\omega,\mathbf{k})$ are the local Green's functions
of the lattice model. 

We solve for all of the lattice and impurity Green's functions with the NESS-DMFT algorithm described in~\cite{AronKotliarWeber2012}. The self-energy kernels $\Sigma_U^{R/K}$ are now computed by solving the following impurity model with the steady-state impurity solver that we describe at the end of this paper. The IPT solution is used as an initial guess.

\begin{figure}[!t]
\centerline{
\includegraphics[height=0.8\columnwidth,angle=-90]{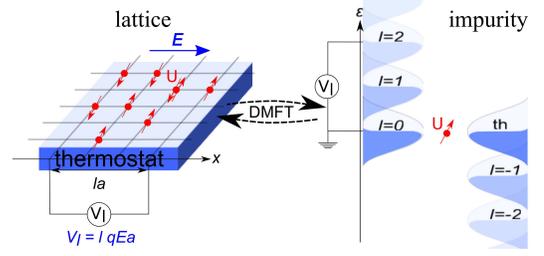}
}
\caption{\label{fig:model}\footnotesize 
The impurity model consists of an impurity coupled to a local thermostat (th) and a set of leads with chemical potential $\mu_l=l \, qEa$, $l\in\mathbb{Z}$. In the framework of NESS-DMFT, the electric-field-driven Hubbard model in contact with a thermostat is solved through a self-consistent mapping to the impurity model.}
\end{figure}

\medskip

\paragraph{\textbf{Impurity model}.} 
Let us re-interpret the Weiss fields as the result of integrating over the degrees of freedom of an effective environment described by the hybridization kernels $\Sigma^{R/K}_{\rm env}$. The out-of-equilibrium nature of the environment is encoded in its energy distribution function
$f_{\rm env}(\epsilon)$  defined by $\Sigma_{\rm env}^K = (2 f_{\rm env} - 1) \, \mbox{Im} \Sigma_{\rm env}^R$. 
The non-interacting Green's functions of the impurity obey the following Schwinger-Dyson equations (assuming that a steady-state can be reached) 
\begin{align}\label{eq:Dyson_env}
\left\{
\begin{array}{rl}
 \mathcal G_0^R(\omega) &= \left[ \omega - \Sigma^R_{\rm env}(\omega) \right]^{-1}\,, \\
 \mathcal G_0^K(\omega) &= \left| \mathcal G_0^R(\omega)\right|^2 \Sigma^K_{\rm env}(\omega)\,.
\end{array}
\right.
\end{align}
Similarly to the equilibrium case, there is no unique way of modeling the impurity environment since different environments can yield the same hybridization kernels. Nonetheless, a model motivated on solid physical ground will facilitate suitable solution schemes. Below, we first propose a generic impurity model, then we specialize it to the case of the electric-field-driven Hubbard model.

Let us represent the non-equilibrium environment as a collection of \textit{equilibrium} non-interacting fermionic reservoirs, labeled by $l$, with different density of states $\rho_l(\epsilon)$, temperatures $T_l$ and chemical potentials $\mu_l$. These reservoirs are linearly coupled to the impurity with the coupling constants $t_l$. The corresponding hybridization kernels read
\begin{align}
\!\!\!
\left\{\label{eq:Sigma_env}
\begin{array}{rl}
 \!\mbox{Im} \Sigma_{\rm env}^R(\omega) \!\! &\!\!= - \pi  \sum_{l} t_l^2 \rho_{l}(\omega) + \mbox{Im}\Sigma_{\rm th}^R(\omega)  \,, \\
 \!\Sigma^K_{\rm env}(\omega) \!\! &\!\!= \pi  \sum_{l} t_l^2 \tanh\left(\frac{\omega-\mu_l}{2 T_l}\right) \rho_{l}(\omega)
 + \Sigma_{\rm th}^K(\omega) 
\,,
\end{array}
\right.\!\!\!
\end{align}
in which we explicitly included the hybridization with the local thermostat.
This representation is fairly generic and can be adapted to a wide class of steady-state impurity problems since it can accommodate any $\Sigma_{\rm env}^R(\omega)$  and $\Sigma_{\rm env}^K(\omega)$.

We now specialize this approach to the case of the electric-field-driven Hubbard model.
Let us consider for a while the potential gauge in which the electric field is rendered by a linear ramp of the scalar potential throughout the lattice: $\phi(t,\mathbf{x})= -q\mathbf{E}\cdot\mathbf{x}$. If one singles out a site, its direct environment is composed of the local thermostat and of its four neighbors: two are on the equi-potential (along the $y$ direction) and the two others are shifted by $\pm qEa$. All sites being equivalent, each of these neighbors feels the same kind of environment, and so on and so forth.

Based on this discussion, we model the impurity environment by the local thermostat and a discrete ``ladder'' of leads shifted in energy by $V_l \equiv l\, \varphi$ with $\varphi \equiv qEa$ and $l\in\mathbb{Z}$. Their chemical potentials are shifted accordingly: $\mu_l = V_l$. A schematic representation of the model is depicted in Fig.~\ref{fig:model}.
In our symmetric case, we have $t_l = t_{-l}$ and we expect $t_0/2 \geq t_1 \geq t_2 \geq ...\,$.

In the absence of a drive ($E=0$), all the leads become degenerate and they can be considered as a unique equilibrium lead at temperature $T$ and zero chemical potential. When the electric field is larger than any other energy scale ($E\to\infty$), the subsequent dimensional reduction~\cite{AronKotliarWeber2012} implies that only the $l=0$ lead is relevant, and that it is in equilibrium at temperature $T$ and chemical potential $\mu_0=0$. 
In the atomic limit ($U\to\infty$), the impurity decouples from the leads, $t_l\to0$, and it equilibrates with the local thermostat.

This impurity model is strictly equivalent to an impurity coupled to equilibrium reservoirs at the same chemical potential (\textit{i.e.} $\mu_l=0$) \textit{via} some time-dependent couplings $t_l \rme^{\rmi V_l t}$. This equivalence corresponds on the lattice side to the $\mathrm{U}(1)$ gauge invariance which states that the electric field can be rendered by a linear ramp potential (potential gauge) or by some time-dependent hopping parameters (so-called temporal gauge).

\begin{figure}[!t]
\vspace{-0.5cm}
\centerline{
\includegraphics[height=0.8\columnwidth,angle=-90]{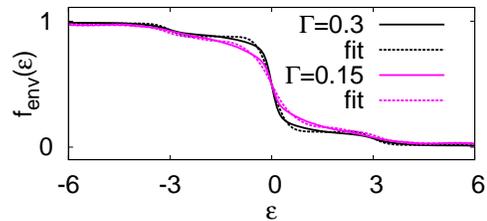}
\vspace{-1cm}
}
\caption{\label{fig:fd_IPT}\footnotesize $f_{\rm env}(\epsilon)$ obtained with the IPT solver is fitted by the expression~(\ref{eq:f_env}) for two values of the dissipation $\Gamma$,  $U=6$ and $E=3$ ($T=0.05$). The lead temperature is found to be $T_L \approx 0.2$ for $\Gamma=0.3$ and $T_L \approx 0.3$ for $\Gamma=0.15$. The hopping parameters are $(t_0,t_1,t_2) \approx (0.38, 0.10, 0.020)$ and $(0.34, 0.13, 0.022)$ respectively, and $t_l \approx 0$ for $l \geq 3$ in both cases.
\vspace{-0.5cm}
}
\end{figure}

We now make the simplifying assumptions that the leads are in equilibrium at the same temperature $T_l=T_L$ and have the same density of states (modulo the energy shift) so that $\rho_l(\epsilon) = \rho_L(\epsilon-V_l)$.
This particular environment yields a steady-state energy distribution function
\begin{align} \label{eq:f_env}
 f_{\rm env}(\epsilon)
 &= \frac{\pi \sum_{l} t_l^2 \rho_L(\epsilon-V_l) f_{T_L}\left({\epsilon-V_l}\right)  + \Gamma f_{T}(\epsilon) }{\pi \sum_{l} t_l^2 \rho_L(\epsilon-V_l) + \Gamma} \,,
\end{align}
where $f_{T_L}(\epsilon)$ is the equilibrium Fermi-Dirac distribution at temperature $T_L$. Notice that although the distribution function $f_{\rm env}$ is clearly not thermal for finite electric fields (see for instance Fig.~\ref{fig:fd_IPT}), $T_L$ sets a new thermal fluctuation scale in the problem.
$T_L$ reduces to $T$ in equilibrium ($E=0$ or $E\to\infty$) and it diverges for a vanishing dissipation ($\Gamma\to0$) as soon as both the drive $E$ and the interaction $U$ are finite because the lattice side overheats in the absence of a dissipation mechanism.

\begin{figure}[!t]
\centerline{
\hspace{-0.5cm}
\includegraphics[height=0.6\columnwidth,angle=-90]{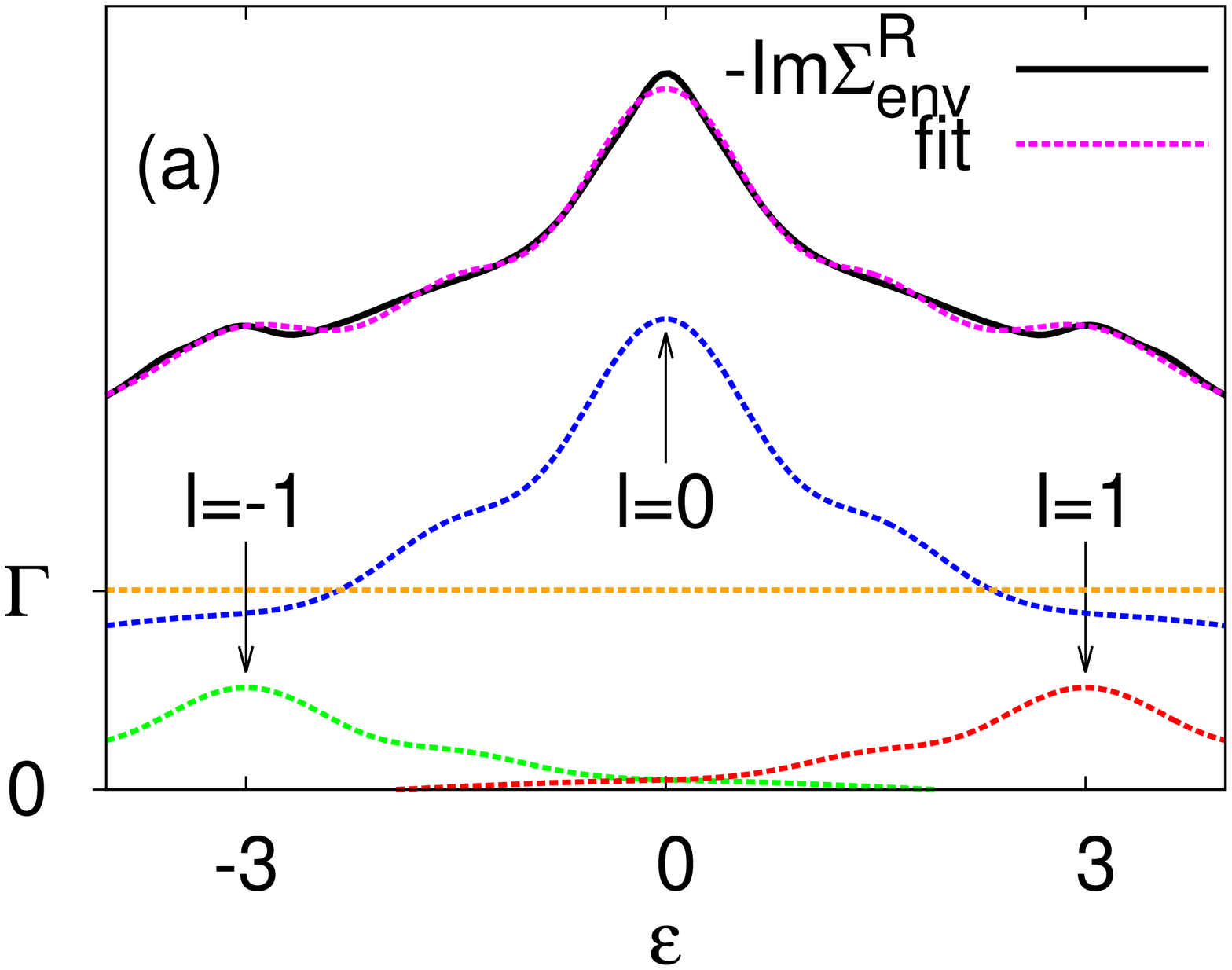}
\hspace{-1cm}
\includegraphics[height=0.6\columnwidth,angle=-90]{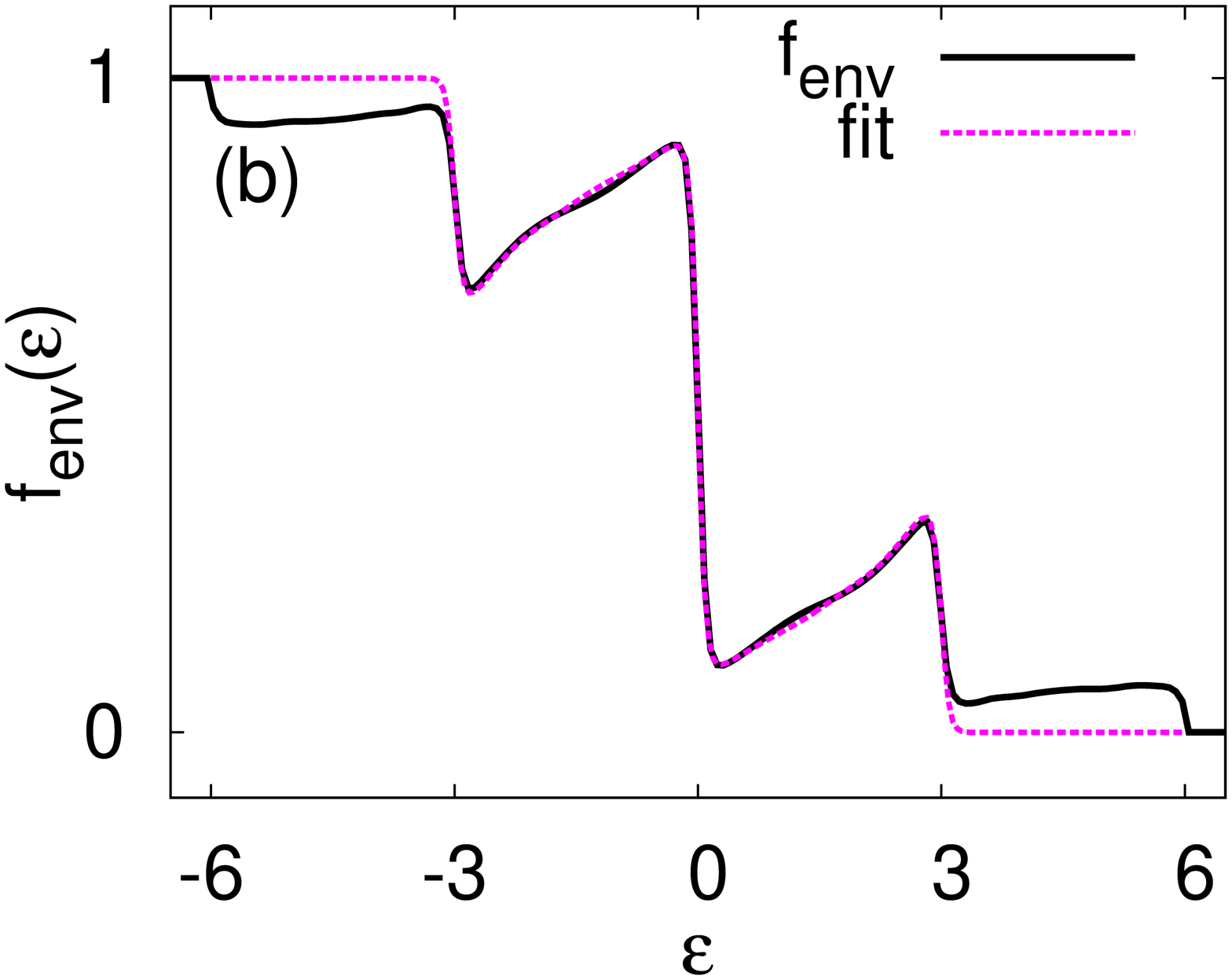}
\hspace{-1cm}
}
\caption{\label{fig:fits}\footnotesize Fit of the environment: joint determination of $\rho_L(\epsilon)$, $T_L$, $t_{0}$, and $t_{1}$ for $U=3$ and $E=3$ ($T=0.05, \Gamma=0.3$).
(a) $\mbox{Im} \Sigma^R_{\rm env}(\epsilon)$ (plain line) is fitted with the dashed line which is the sum of the terms $l=-1$ (green), $l=0$ (blue), $l=1$ (red) and $\Gamma$ (orange) in Eq.~(\ref{eq:Sigma_env}).
(b) $f_{\rm env}(\epsilon)$ (plain line) is fitted with the dashed line by the expression~(\ref{eq:f_env}). 
The other parameters of the fit are $T_L \approx 0.05$, $t_0 \approx 0.32$ and $t_1 \approx 0.18$.
}
\end{figure}

We validate the model by demonstrating that the right-hand sides of Eqs.~(\ref{eq:Sigma_env}) and (\ref{eq:f_env}) can successfully reproduce their left-hand sides that are generated by the NESS-DMFT algorithm. In practice, we use a Newton's steepest descent method to fit the parameters of the model (the $t_l$'s, $T_L$ and $\rho_L$). In Fig.~\ref{fig:fd_IPT}, we provide some examples of this fitting procedure on the DMFT solutions obtained with the IPT solver for $U=6$, $E=3$. At this value of the interaction, the Hubbard bands are already present in the spectrum and the quasi-particle peak melts down when $qEa \sim U/2$~\cite{Aron2012}. Figure~\ref{fig:fd_IPT} shows that this melt-down translates on the impurity side into a high lead temperature $T_L \gg T$ which increases as $\Gamma$ decreases.
We now discuss the case in which the DMFT solutions are obtained with the non-perturbative steady-state impurity solver that is to be introduced below. Given the limitations of this solver, we work with parameters $E$ and $U$ such that (i) $T \simeq T_L$, and (ii) we can truncate the model to a small number of leads, $l=-1, 0, 1$, and discard the other leads~\footnote{This approximation is valid as long as $t_0^2 + t_1^2 \gg \sum_{l=2}^{\infty} t_l^2$.}. 
In Fig.~\ref{fig:fits}(a), we give the results of the fitting procedure for $U=3$ and $E=3$ by plotting $\mbox{Im}\Sigma_{\rm env}^R(\epsilon)$ as the sum of $t_{1}^2 \rho_{L}(\epsilon-\varphi)$, $t_{0}^2 \rho_{L}(\epsilon)$, $t_{1}^2 \rho_{L}(\epsilon+\varphi)$ and $\Gamma$. The local maxima of $\mbox{Im}\Sigma_{\rm env}(\epsilon)$ at $\epsilon=\pm qEa$ and $\epsilon =0$ naturally come from the replication of the unique maximum of $\rho_L(\epsilon)$ at $\epsilon=0$.
In Fig.~\ref{fig:fits}(b), we plot the corresponding energy distribution function $f_{\rm env}(\epsilon)$. The jump around $\epsilon=0$ is accounted for by the $l=0$ lead and the thermostat while the two symmetric jumps around $\epsilon=\pm qEa$ are described by the leads $l=\pm 1$. As a result of the truncation, the high-energy features above $|\epsilon| > |q|Ea$ are not properly captured and $t_1$ is slightly overshot.

\medskip

\paragraph{\textbf{Impurity solver}.}
To demonstrate the practical relevance of the impurity model we introduced, we solve it and compute non-perturbatively the interaction contribution to the retarded self-energies $\Sigma_U^R$ by generalizing to multiple leads the steady-state impurity solver that was recently developed by Han and Heary in the context of a two-lead environment~\cite{Han-Heary, PruschkeHan}. Unlike the IPT solver, this solver provides conserving solutions even away from particle-hole symmetry.

Building on Hershfield's expression for the steady-state density matrix~\cite{Hershfield}, Han and Heary gave an effective Matsubara description of the steady state. This allows the use of standard many-body equilibrium tools to tackle the strong interaction at the cost of introducing some imaginary chemical potentials: $\rmi\varphi_m \equiv \rmi\, {2m\pi} \, T_L$, $m\in\mathbb{Z}$. After the impurity problem is solved in imaginary time for each $m$, the solutions are analytically continued to the real-time real-voltage domain.

We refer the reader to~\cite{Han} for the technical details as we follow closely the steps explained there. The non-interacting impurity Green's function in Matsubara frequency and imaginary-voltage reads:
\begin{equation}
 {\cal G}_0(\rmi\omega_n,\rmi\varphi_m) = \!\!\!\!\sum_{\!l=0,\pm1} \! \int\Ud{\epsilon}\!  \frac{\pi t_l^2 \rho_L(\epsilon-V_l) +  \Gamma \delta_{l0}}{\pi \sum_{p}  \! t_p^2 \rho_L(\epsilon-V_p) + \Gamma }  \frac{\rho_0(\epsilon)}{z_{nm,l}-\epsilon}\,,
\end{equation}
with $z_{nm,l}\equiv\rmi\omega_n-l({\rmi\varphi_m-\varphi})$, $\rmi\omega_n \equiv \rmi(2n+1) \pi T_L$, $\rho_0(\epsilon) \equiv -\mbox{Im} \mathcal{G}^R_0(\epsilon)/\pi$ and $\delta_{l0}=1$ if $l=0$ and 0 otherwise. The differences with Eq.~(32) in~\cite{Han} lie in the presence of the $l=0$ term and the frequency-dependent environment which is determined self-consistently.

\begin{figure}[!t]
\hspace{-0.8cm}
\includegraphics[height=0.6\columnwidth,angle=-90]{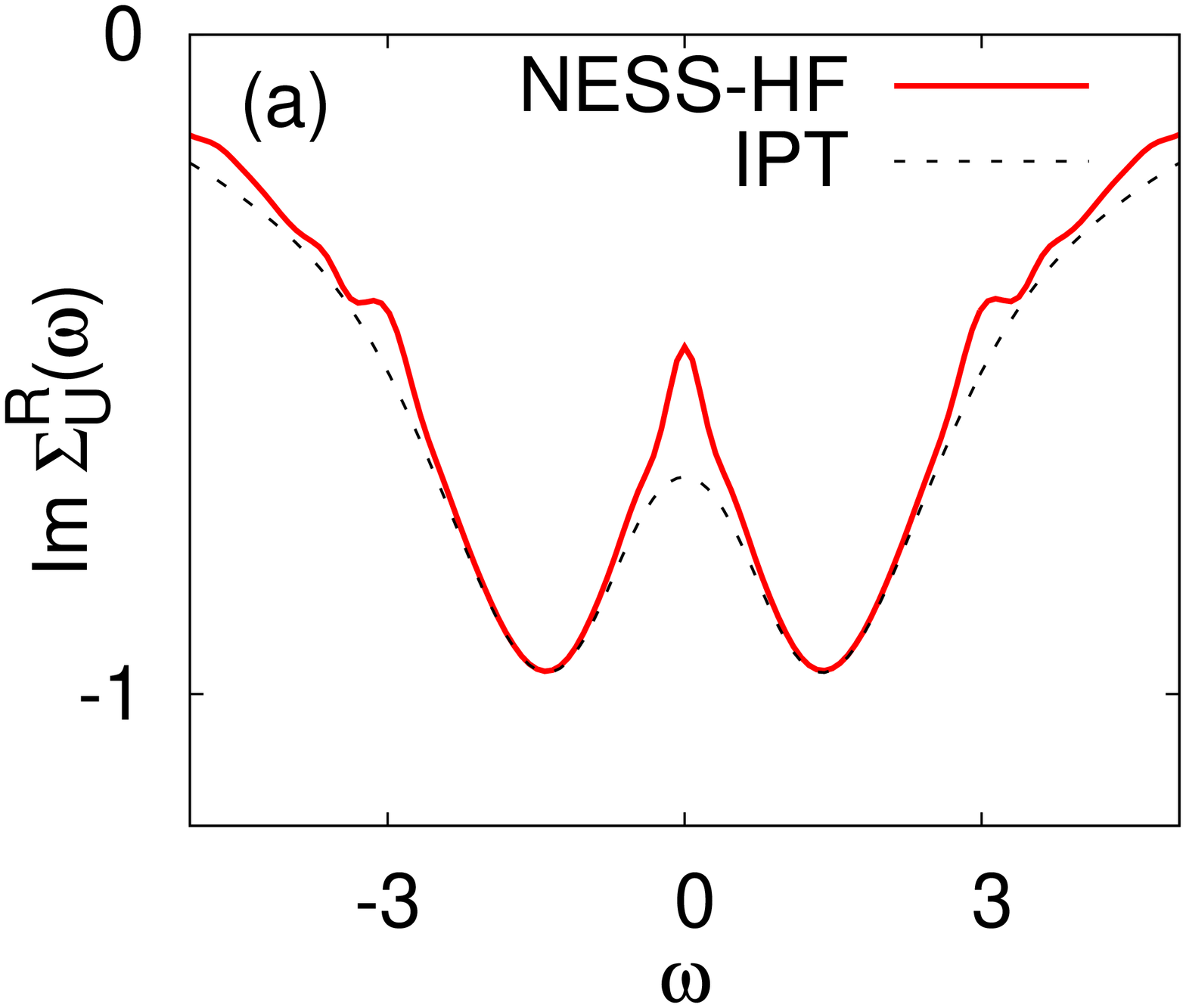}
\hspace{-1.2cm}
\includegraphics[height=0.6\columnwidth,angle=-90]{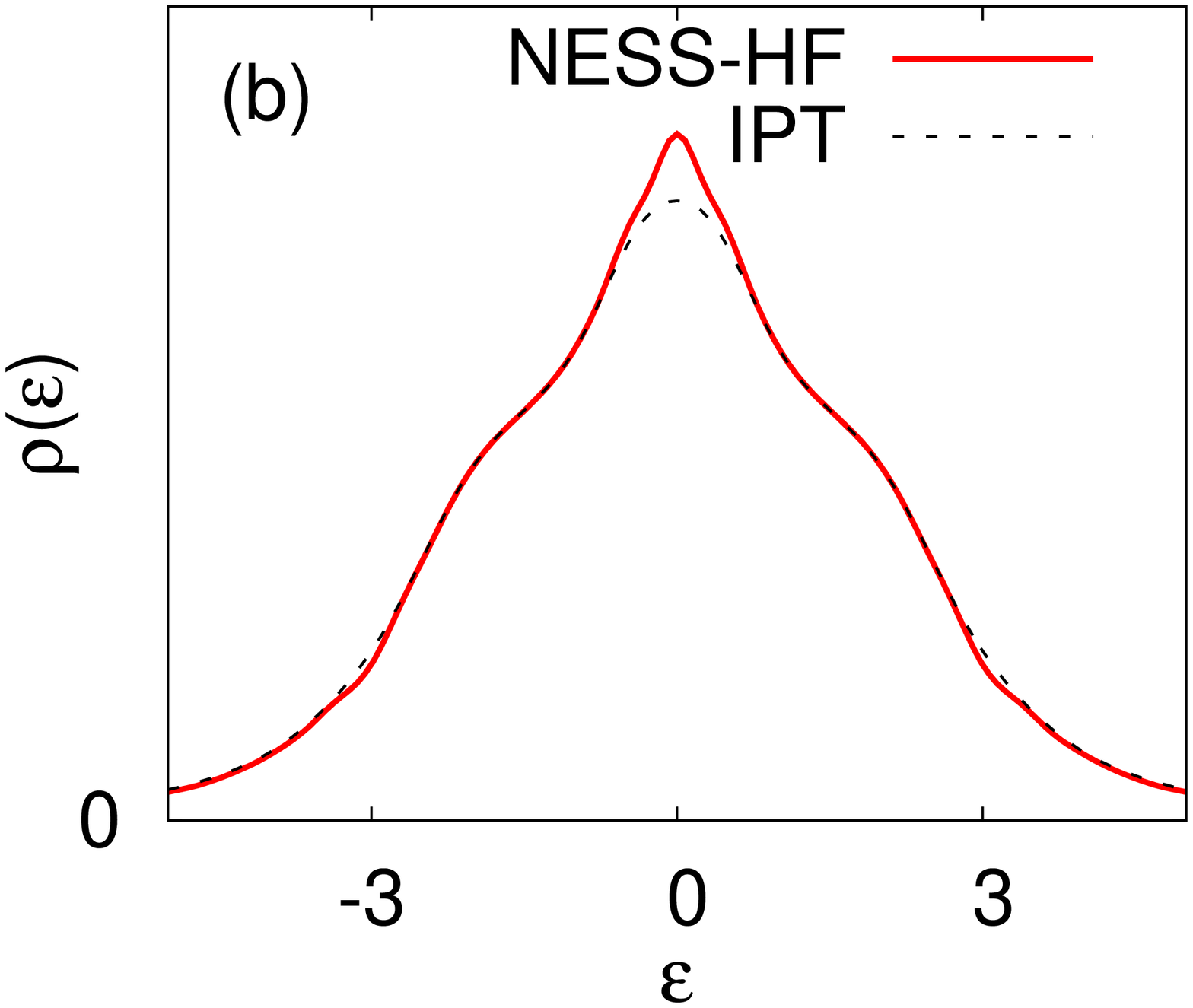}
\hspace{-1cm}
\caption{\label{fig:Sigma_U_R}\footnotesize (a) Retarded self-energy $\mbox{Im}\Sigma_U^R(\omega)$ obtained by the NESS-QMC solver. (b) Local density of states $\rho(\epsilon)$. The dashed lines correspond to the IPT solution. ($U=3$, $E=3$, $T=0.05$, $\Gamma=0.3$).}
\end{figure}

The retarded self-energies $\Sigma_U^R(\rmi\omega_n,\rmi\varphi_m)$ are obtained by means of a Hirsch-Fye algorithm~\cite{HF}. More sophisticated algorithms have already been used in the context of a two-lead static environment~\cite{Pruschke-Werner}.
$\Sigma_U^R(\omega)$ is then obtained after a double analytical continuation: $\rmi\omega_n \to \omega$ and $\rmi\varphi_m \to \varphi$. This is performed numerically by fitting all the $\Sigma_U^R(\rmi\omega_n,\rmi\varphi_m)$ by the following ansatz which is guessed from the expression of the self-energy to the second order in $U$, and which has been validated in our regime of parameters~\cite{Han}:
\begin{align}
 \Sigma_U^R(\rmi\omega_n,\rmi\varphi_m) = \sum_{\gamma\in\mathbb{Z}} \int \ud{\epsilon} \frac{\sigma_\gamma(\epsilon) Q_\gamma(\epsilon,\rmi\varphi_m-\varphi)}{z_{nm,\gamma}  - \epsilon}\,.
\end{align}
We truncate the sum at the order $\gamma_{\rm max} = 5$ and the function $Q_\gamma(\epsilon,z)$ is fitted by the simple Pad\'e approximant $\frac{1+C_\gamma(\epsilon)z}{1+D_\gamma(\epsilon)z}$. The parameters of the fit are the functions $\sigma_\gamma$, $C_\gamma$ and $D_\gamma$. The real-frequency retarded self-energy is obtained by
\begin{align}
 \mbox{Im} \Sigma_U^R(\omega) = -\pi \sum_\gamma \sigma_\gamma(\omega)\,.
\end{align}
We used a minimization procedure based on Newton's steepest descent method.
Although the success of analytical continuations is never exempt from some numerical `cooking' techniques, this double analytical continuation was unexpectedly rather easy to perform, perhaps thanks to the presence of a finite dissipation in the bulk that moves dangerous poles away from the real axis. In Fig.~\ref{fig:Sigma_U_R}(a) we plot $\mbox{Im}\Sigma_U^R(\omega)$ for $U=3$ and $E=3$ and compare with the result obtained with the IPT solver. The sharper edges around $\omega=0$ are responsible for a better defined quasi-particle peak in the local density of states [see Fig.~\ref{fig:Sigma_U_R}(b)] but altogether, the results are quite close to the ones obtained with IPT. This agreement for relatively small $U$ validates our impurity model and impurity solver.

We approximate the steady-state distribution function of the interacting impurity by the one of its environment: $f(\epsilon) = f_{\rm env}(\epsilon)$. 
It is a reasonable approximation for the following reasons:
(i) it is exact in equilibrium ($E=0$ and $E\to\infty$), 
(ii) it is exact in the non-interacting case ($U=0$) and in the atomic limit ($U\to\infty$),
and (iii) it is consistent with charge and energy conservations.
In turn, this gives us a simple way to estimate the Keldysh component of the self-energy:
\begin{align}
\Sigma_U^K(\omega) = [2f(\omega)-1]\, \mbox{Im} \Sigma_U^R(\omega)\;.
\end{align}
Once the impurity is solved, the self-energy kernels $\Sigma_U^{R/K}$ are used to recompute the lattice Green's functions in Eqs.~(\ref{eq:DysonLattice}), a new impurity problem is defined and solved again, so on and so forth until convergence is reached.

\medskip

\paragraph{\textbf{Conclusion}.}
Although the work presented here is centered around the electric-field-driven Hubbard model, the impurity model, the dictionary between the lattice side and the impurity side, and the steady-state impurity solver allow to consider further questions, such as the effect of chemical substitution (doping) and pressure effects under an electric field, not achievable with simpler perturbation methods. 
Also, not only does the steady-state impurity solver enable to study the transport properties of quantum dots driven by leads with a realistic density of states, it can also address fundamental questions such as the effect of a quantum critical point (gapped leads) on the out-of-equilibrium Kondo physics~\cite{Schiro}.

We are grateful to N. Andrei, P. Dutt, J.E. Han, K. Le Hur, and O. Parcollet for comments and discussions. This work has been supported by NSF grant No. DMR-0906943.

\end{document}